\documentclass[11pt,twoside]{article}

\usepackage{asp2006}
\usepackage{epsf}
\usepackage{psfig}
\usepackage{lscape}

\def\kms      {\ifmmode{\rm km\,s}^{-1} \else km\,s$^{-1}$\fi}
\def\mujybm{\ifmmode{\rm \mu Jy}\,{\rm beam}^{-1}\else${\rm \mu}$Jy\,beam$^{-1}$\fi}
\def\ltsim{\ifmmode\stackrel{<}{_{\sim}}\else$\stackrel{<}{_{\sim}}$\fi}
\def\gtsim{\ifmmode\stackrel{>}{_{\sim}}\else$\stackrel{>}{_{\sim}}$\fi}
\def\farcs{\hbox{$.\!\!^{\prime\prime}$}}
\def\farms{\hbox{$.\!\!^{\prime}$}}
\def\fsec{\hbox{$.\!^{\rm s}$}}
\def\hour{\hbox{$^{\rm h}$}}
\def\min{\hbox{$^{\rm m}$}}
\def\mum{$\mu$m}
\def\spitzer{{\it Spitzer}}

\def\degr{\hbox{$^{\rm o}$}}

\begin{document}
\title{Sub-arcsecond, microJy radio properties of \spitzer\ identified
mid-infrared sources in the HDF-N/GOODS-N field}   %%% Fill in title
\author{R. J. Beswick, T. W. B. Muxlow, H. Thrall, A. M. S. Richards}   %%% Fill in author names
\affil{Jodrell Bank Observatory, The University of Manchester, Lower
Withington, Macclesfield, SK11~9DL, United Kingdom}    %%% Fill in author affiliations

\begin{abstract} %%% Abstract to run on from here.

We present recent and ongoing results from extremely deep
18 day MERLIN + VLA 1.4GHz observations (rms: 3.3microJy/bm) of an 8.5$\times$8.5 arcminute field centred upon the Hubble Deep Field North. This area of
sky has been the subject of some of the deepest observations ever made
over a wide range of frequencies, from X-rays to the radio. The
results presented here use our deep, sub-arcsecond radio imaging of this field to characterise the
radio structures of the several hundred GOODS {\it Spitzer} MIR
sources   in this field. These MIR sources primarily trace the luminous starburst sources. A
significant proportion of the MIR sources are detected and resolved by
our radio observations, allowing these observations to trace the
IR/Radio correlation for galaxies over $\sim$7 orders of magnitude,
extending it to ever lower luminosities.
\end{abstract}

\section{Introduction}

Over the last 3 decades studies of the radio and far-infrared (FIR)
properties of galaxies have shown there to be a tight
correlation between the emission from galaxies in these two observing
bands that extends over several orders of magnitude in luminosity
\citep{vanderkruit73,condon82}.  More recently investigations using
the mid-infrared (MIR) bands at 24\,$\mu$m and 70\,$\mu$m of {\it Spitzer}  
by \cite{appleton04}, have shown that the  correlation of the MIR and FIR
emission to the radio emission persist out to redshifts of at least 1,
with deep field observations by \cite{garrett02} tentatively extending
this correlation out to $z\sim$4.  The correlation between the radio
and infrared emission arises because both are related to the star
formation processes; the infrared emission is produced from dust
heated by photons from young stars, while the radio emission arises from
synchrotron radiation produced by the acceleration of charged
particles from supernovae explosions.  Investigations of nearby
star-forming galaxies \citep{murphy06} have begun to extend this
correlations to even fainter
luminosities (down to L$_{24\,\mu{\rm m}}\approx10^{20}$\,W\,Hz$^{-1}$) by
considering discrete regions within individual galaxies. However,
studies of deep field at radio and IR wavelengths still uniquely
provide a method by which this correlation can be investigated using many faint galaxies
out to high redshifts.

Deep radio observations of the HDF-N region were made in
1996-97 at 1.4\,GHz using both MERLIN and the VLA. The initial results
of these observation were presented in \cite{muxlow05},
\cite{richards98} and \cite{richards00}. Within these radio observations
92 sources were detected brighter than a completeness limit of
40\,$\mu$Jy\,beam$^{-1}$ (5.3$\sigma$) within
a 10 square arcminute field, using a 2\arcsec\ beam. \cite{muxlow05} presented results from the
combination of both this 42\,hr VLA observation and an 18 day MERLIN
integration at the same pointing centre
($\alpha=12$\hour\,36\min\,49\fsec4000,
$\delta=+62$\degr\,12\arcmin\,58\farcs000 (J2000)). The combined
MERLIN$+$VLA image has an rms noise level of 3.3\,$\mu$Jy per 0\farcs2 circular
beam making it amongst the most sensitive high resolution radio image
made to date.  Full details of these observations and data reduction
are presented in \cite{muxlow05} and Muxlow {\it et al.} ({\it this proceedings}).
\begin{figure}
\begin{center}
\setlength{\unitlength}{1mm}
\begin{picture}(75,55)
\put(0,0){\includegraphics{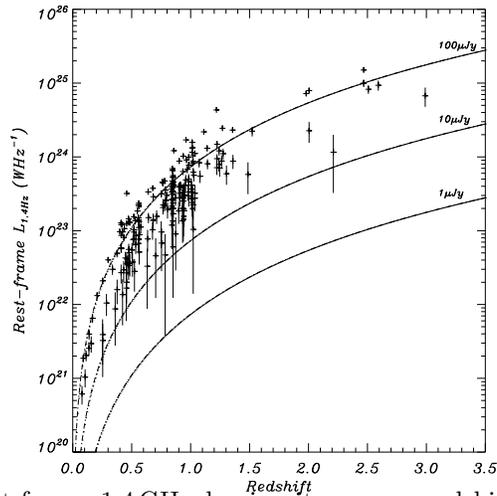}}
\end{picture}
\caption{Rest-frame 1.4\,GHz luminosity versus redshift for 212
24\,\mum\ \spitzer\ sources with redshifts and 1.4\,GHz flux densities
greater than 3 times the local map rms. All 1.4\,GHz luminosities have measured within an
aperture of radius 1\farcs5 centred upon the \spitzer\ MIPS 24\,\mum\ source
position.}
\label{Beslumvsz}       % Give a unique label
\end{center}
\end{figure}

In this proceedings we present some preliminary results from an
extended analysis of these deep MERLIN$+$VLA radio observations of the
HDF-N. This extended analysis uses the ancillary data obtained from
the large multiwavelength Great Observatories Origins Survey (GOODS)
which when combined with these extremely deep radio observations
provide a clear insight into the characteristics of the microJansky
radio source population. In these proceedings we briefly outline some
of these preliminary results with particular emphasis on the radio
detected {\it Spitzer} 24\,\mum\ sources within the 10$\times$10 arcminute radio field. Additional further analysis of these radio
observations, exploiting GOODS data-sets at X-ray and optical
wavelengths are also presented in these proceedings by Richards {\it
et al}, Thrall {\it et al} and Muxlow {\it et al.}. The results
presented here are covered in more much detail in
Beswick {\it et al. (in prep)}.   

\begin{figure*}
\begin{center}
\setlength{\unitlength}{1mm}
\begin{picture}(75,45)
\put(0,0){\includegraphics{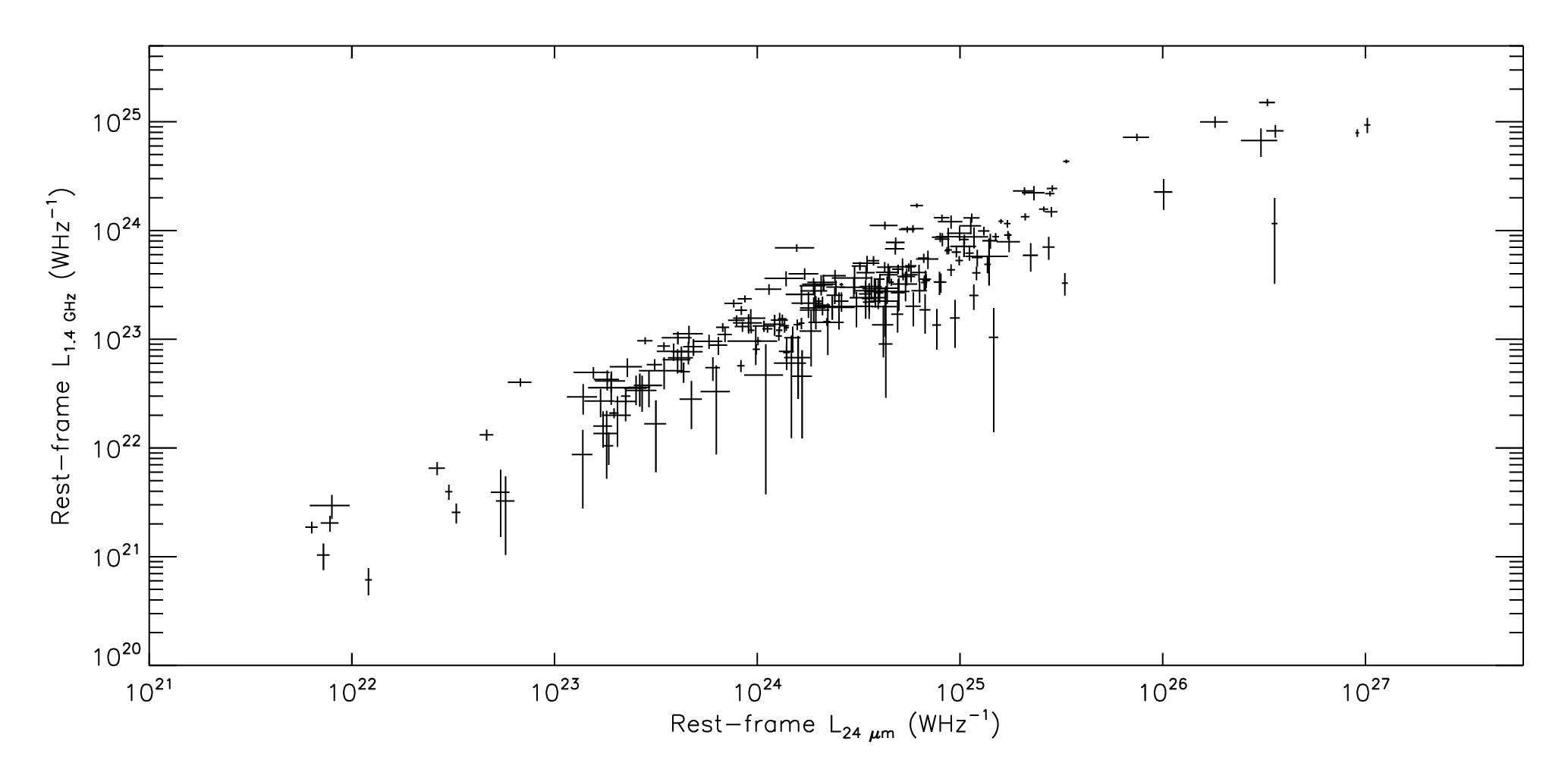}}
\end{picture}
\caption{The rest-frame luminosities at 1.4\,GHz and 24\,\mum\ for 212
24\,\mum\ \spitzer\ sources with redshifts. The rest-frame 1.4\,GHz luminosities
have been k-corrected (boosted) assuming a spectral index of 0.7. The
rest-frame 24\,\mum\ luminosities have been corrected assuming an
'Arp220-like' SED. All errors shown are equivalent to
3-$\sigma$. All 1.4\,GHz luminosities have measured within an
aperture of radius 1\farcs5 centred upon the \spitzer\ MIPS 24\,\mum\ source
position. Note that in the many of cases where sources with apparently
low radio luminosity with respect to MIR luminosity are large nearby
spirals with an optical extent larger than the 3\arcsec.}
\label{Beslumlum}       % Give a unique label 
\end{center}
\end{figure*}

\section{Results \& Discussion}

\subsection{Radio luminosities of 24\,\mum\ \spitzer\ sources}
Within the entire field covered by the GOODS-N {\it Spitzer} 24\,\mum\ observations
1199 sources were identified with flux densities
$>$80\,$\mu$Jy. However, for the purposes of comparing these
24\,\mum\ source with our radio data we have limited this sample to
sources that are
spatially coincident with the 8.5$\times$8.5 arcmin$^2$ radio field. This
sub-sample contains 377 {\it Spitzer} sources with 24\,\mum\  flux
densities ranging from 80.1 to 1480\,$\mu$Jy. 

Using the {\`a} prior positional information from the 24\,\mum\
\spitzer\ source catalogue,  radio emission
in our deep 8\farms5$\times$8\farms5 MERLIN$+$VLA
1.4\,GHz image was searched for within a series of eight,
evenly-spaced, concentric rings with radii between 0.5 and 4
arcsec, at the position of each of the 377 24\,\mum\ sources. From the
statistical analysis of the flux densities recorded within each of
these annuli it has been shown that the majority ($>$90$\%$) of the radio
flux density of these sources is recovered within a radius of 1\farcs5
of the \spitzer\ source positions (Beswick {it et al. in prep}). A plot of the 1.4\,GHz radio
luminosity of all of the 24\,\mum\ sources with known redshift against
their redshift is shown in Fig.\,\ref{Beslumvsz}. As can be seen in
Fig.\,\ref{Beslumvsz}, the majority of the \spitzer\ 
sources lie between redshifts 0 and 3.5, and have 1.4\,GHz radio flux
densities of less than 100\,$\mu$Jy. From the radio morphological and
spectral arguments of \cite{muxlow05} it has been shown that the
faint radio source population (below 100\,$\mu$Jy) is dominated by
star-forming galaxies. Similarly it can be expected that this MIR
selected sample of radio faint intermediate and high redshift sources
also primarily consists of star-forming galaxies. The \spitzer\
24\,\mum\ versus 1.4\,GHz luminosities of these sources is plotted in
Fig.\,\ref{Beslumlum}. In this figure it can be seen that as expected
the MIR and radio luminosities of these star-forming
galaxies are strongly correlated over $\sim$7 orders of magnitude:
extending the correlation by several orders of magnitude to lower
luminosities.

\subsection{Radio sizes of 24\,\mum\ \spitzer\ sources}

\begin{figure}
\begin{center}
\setlength{\unitlength}{1mm}
\begin{picture}(75,40)
\put(0,0){\includegraphics{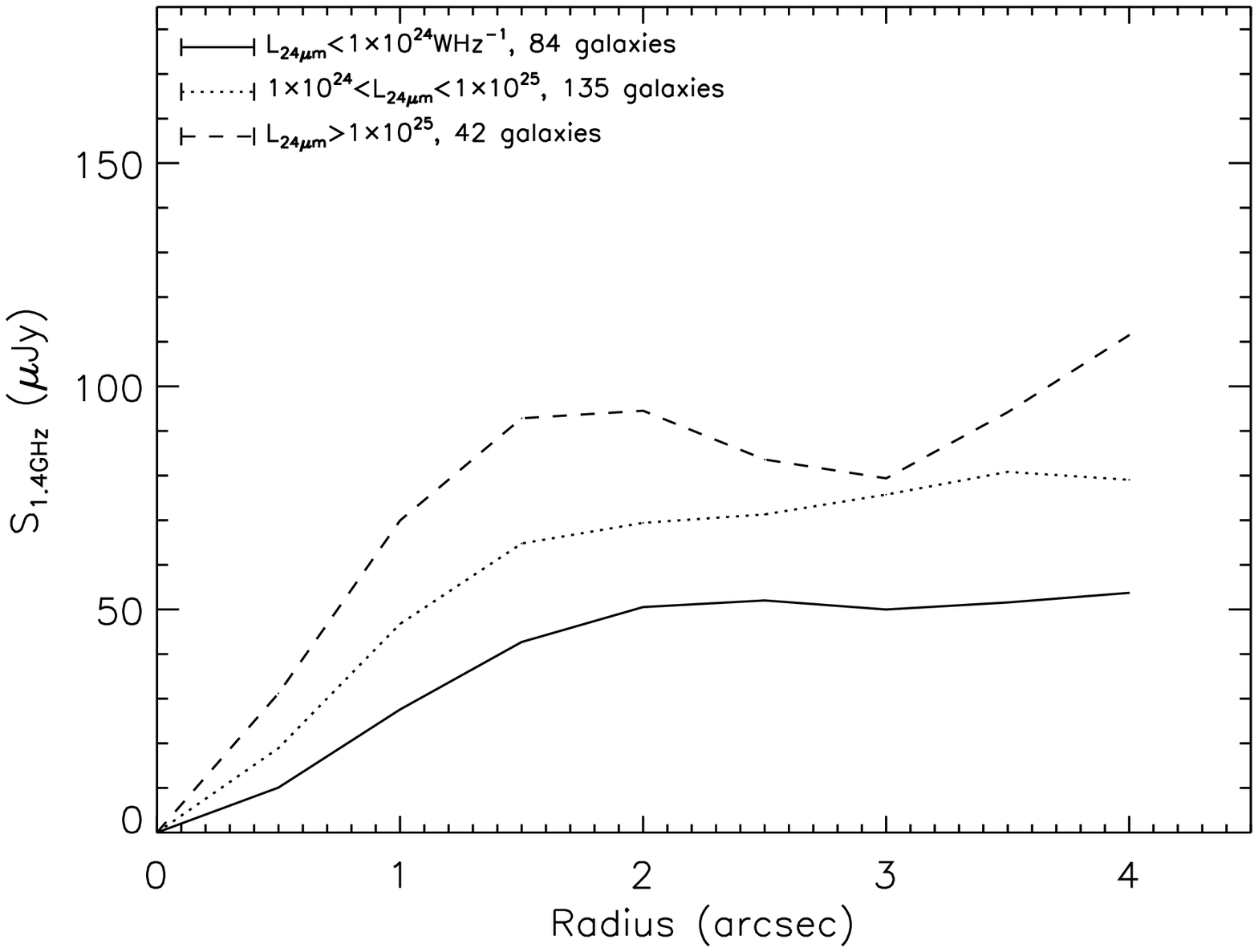}}
\put(0,0){\includegraphics{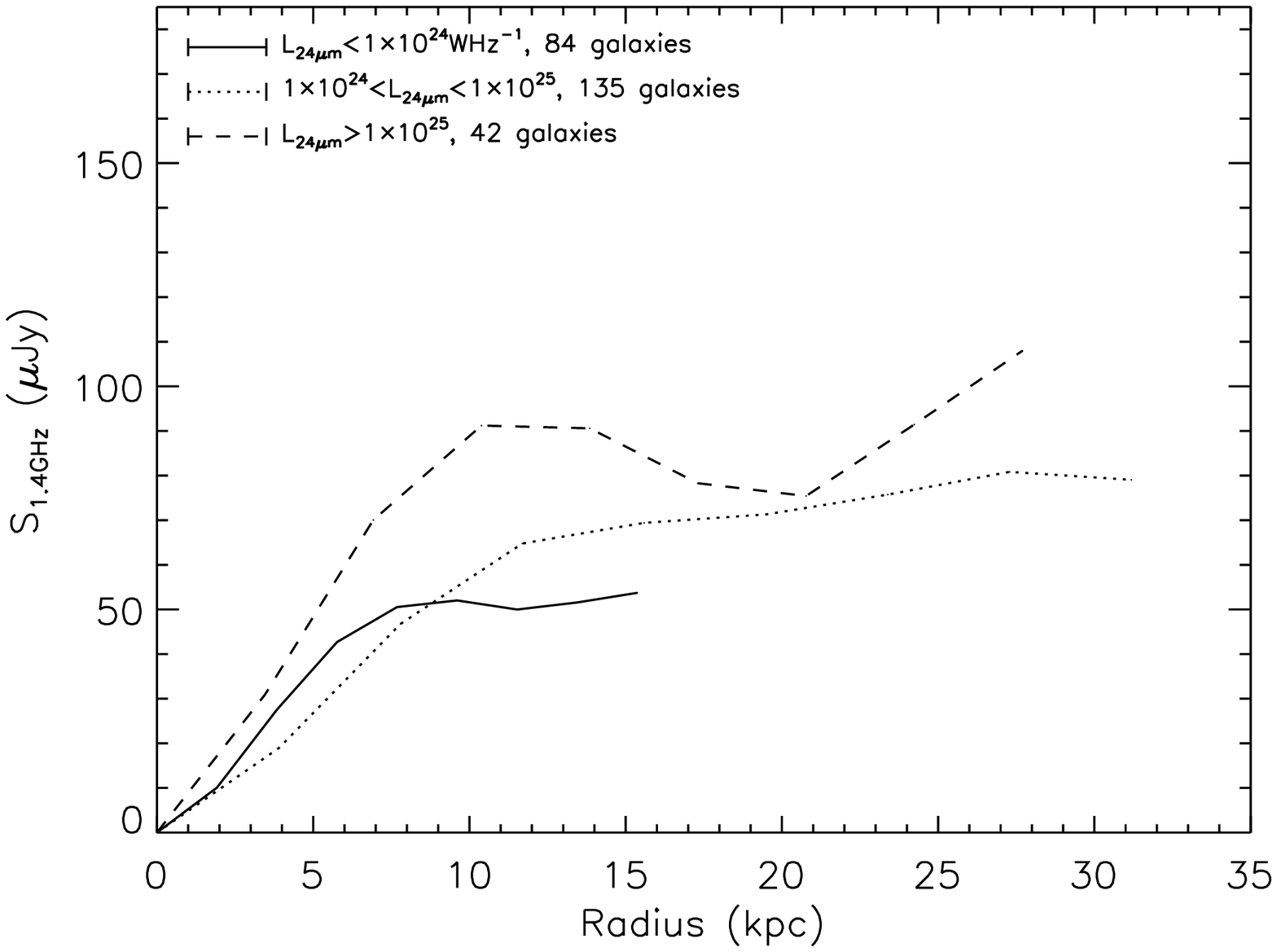}}
\end{picture}
\caption{Radial plots of the mean cumulative enclosed 1.4\,GHz flux density versus
radius for the 261
24\,\mum\ \spitzer\ sources with redshifts. The averaged
source flux densities within enclosed radii (plotted against angular and linear
sizes) are shown for sub-sets of
these sources binned with respect to their 24\,\mum\ luminosity. Each
sub-set has been ascribed the mean redshift sources in each bin.}
\label{Bessize-l24}       % Give a unique label
\end{center}
\end{figure}

One massive advantage of these high resolution MERLIN+VLA radio
observations is their sub-arcsecond angular resolution (0\farcs2
$\rightarrow$ 0\farcs5). At a redshift of $\sim$0.7 the resolution of
these radio observations is a few kpc. This angular resolution is adequate to resolve
the galactic scale  radio structure of sources out to redshifts of a
few. Figure \ref{Bessize-l24} shows the cumulative radial profiles of the
1.4\,GHz radio emission of the 24\,\mum\ \spitzer\ sources within the
8\farms5$\times$8\farms5 field binned by their 24\,\mum\
luminosities. As can be seen in the left-hand plot the majority of the
radio flux density for all of these sources is enclosed with in a
radius of 1.5--2 arcseconds. Converting these angular sizes to the
galaxies linear extent (Fig\,\ref{Bessize-l24} {\it right-hand plot})
shows that, on average, the lower luminosity MIR sources (solid line)
have radio emission on sub-galactic scales, extending out to radii of
$\sim$7\,kpc, consistent with radio starburst emission from within these
galaxies. In comparison the radio emission from the more luminous
MIR sources, whilst still on galactic scales, appears to extend to larger radii ($\sim$10-15\,kpc).

\acknowledgements %%% Text of acknowledgements runs on after this command.

RJB acknowledges financial support by the European Commission's I3
Programme ``RADIONET'' under contract No. 505818. HT acknowledges
support from a PPARC studentship. Based on
observations made with MERLIN, a National Facility operated by the
University of Manchester at Jodrell Bank Observatory on behalf of
PPARC, and the VLA of the National Radio Astronomy Observatory is a
facility of the National Science Foundation operated under cooperative
agreement by Associated Universities, Inc.

%%% THE BIBLIOGRAPHY
%%%
%%% CONSULT SECTION 3 OF "INSTRUCTIONS FOR AUTHORS" FOR HOW TO USE NATBIB.
%%% AUTHORS ARE ENCOURAGED TO USE EITHER THE "THEBIBLIOGRAPY" ENVIRONMENT
%%% BY UNCOMMENTING (DELETING THE "%" SYMBOL) THE COMMANDS BELOW, OR BY
%%% USING THE BIBTEX ENVIRONMENT. TO FIND OUT WHICH IS APPLICABLE TO YOUR
%%% CONTRIBUTION, CONSULT THE VOLUME EDITORS FOR YOUR PROCEEDINGS.
%%%

\end{document}